\documentclass[twocolumn,prd,aps,superscriptaddress,showpacs,
nofootinbib,preprintnumbers,floats,floatfix]{revtex4}

\usepackage{amsmath}
\usepackage{amssymb}
\usepackage{latexsym}
\usepackage{graphicx}
\usepackage{hyperref}
\usepackage{bm}

\begin{document}

\title{On what scale should inflationary observables be constrained?}

\author{Marina Cort\^{e}s}
\affiliation{Astronomy Centre, University of Sussex, Brighton BN1
9QH, United Kingdom}
\affiliation{Cosmology and Gravity Group, Department of Mathematics
and Applied Mathematics, University of Cape Town, Cape Town, South
Africa} 
\author{Andrew R. Liddle}
\affiliation{Astronomy Centre, University of Sussex, Brighton BN1
9QH, United Kingdom}
\author{Pia Mukherjee}
\affiliation{Astronomy Centre, University of Sussex, Brighton BN1
9QH, United Kingdom}
\date{\today}

\begin{abstract}
We examine the choice of scale at which constraints on inflationary
observables are presented. We describe an implementation of the
hierarchy of inflationary consistency equations which ensures that
they remain enforced on different scales, and then seek to optimize
the scale for presentation of constraints on marginalized inflationary
parameters from WMAP3 data.  For models with spectral index running,
we find a strong variation of the constraints through the range of
observational scales available, and optimize by finding the scale
which decorrelates constraints on the spectral index $n_{{\rm S}}$ and
the running.  This scale is $k=0.017 \, {\rm Mpc}^{-1}$,
and gives a reduction by a factor of more than four in the allowed
parameter area in the $n_{\rm S}$--$r$ plane ($r$ being the
tensor-to-scalar ratio) relative to $k=0.002 \, {\rm Mpc}^{-1}$. These
optimized constraints are similar to those obtained in the no-running
case. We also extend the analysis to a larger compilation of data,
finding essentially the same conclusions.
\end{abstract}

\pacs{98.80.Cq}
\preprint{astro-ph/0702170}

\maketitle

\section{Introduction}

In presenting observational constraints on the primordial power
spectra, such as those that may have been generated by inflation, it
is necessary to specify the scale at which the observables are being
determined. Provided the full posterior distribution over all
parameters is given, this choice is an arbitrary one. However, if the
information is to be compressed via marginalization, the choice of
this scale matters, and should be chosen in order to optimize the
presentation of constraints.

In the Wilkinson Microwave Anisotropy Probe (WMAP) three-year
cosmological parameters paper \cite{wmap3} the scale $0.002 \, {\rm
Mpc}^{-1}$ is used, which is close to the observable horizon, while
Kurki-Suonio et al.~\cite{KMV} and Finelli et al.~\cite{finelli} found
that the choice of $0.01 \, {\rm Mpc}^{-1}$ worked better in
constraining inflationary observables as it is closer to the
statistical center of the data. The scale $0.05 \, {\rm Mpc}^{-1}$ is
also commonly used, being the default scale of the CosmoMC package
\cite{CosmoMC}.  The pivot scale was also discussed in
Ref.~\cite{LLMP}, who sought the scale where the perturbation
amplitude was best determined (decorrelated with other power spectrum
parameters), and in Ref.~\cite{PE} who sought the scale at which the
perturbation spectrum reconstructed using the flow formalism was best
constrained. 

In this paper we make a systematic exploration of the choice of scale
in the context of inflation models. This choice is particularly
important in cases where models with significant spectral index
running are allowed. Such models have received quite a bit of
attention since the WMAP results emerged (see
e.g.~Refs.~\cite{run,EP,KKMR}).

\section{Methodology}

For definiteness we concentrate on single-field inflationary models,
though many of the issues we discuss are more general. These models
predict spectra of scalar and tensor perturbations which are related
by a hierarchy of consistency equations \cite{LLKCBA,CL}, the first of
which is, at lowest-order, the well-known relation $r=-8n_{{\rm T}}$
where $r$ is the tensor-to-scalar ratio and $n_{{\rm T}}$ the tensor
spectral index. These parameters can in turn be related to the
inflationary slow-roll parameters describing the shape of the
potential.

Our main aim in this paper is to examine the optimal choice of scale
at which to present observational constraints on inflation. In order
to fit the spectra from data, they must first be parametrized, which
is usually done by specifying their amplitude and some number of
derivatives (i.e.~the spectral index, running, etc) at a particular
scale. So far, this scale has been chosen by hand.

The choice of scale, being arbitrary, ought not to affect the
conclusions one draws. However this is only the case if one specifies
the full multi-dimensional posterior parameter distributions, and
provided the model definition is internally self-consistent. The first
of these is often not the case, as one commonly wishes to condense
information into one- or two-dimensional marginalized posterior
distributions, which throws away the information on parameter
correlations necessary to translate between scales. The second
condition of model self-consistency holds in most circumstances, but
often \emph{not} in the way inflationary spectra are implemented.

The problem of model definition in inflationary models is the
enforcement of the consistency equations between scalars and
tensors. Typically both spectra are allowed to be power-laws but with
different indices; if the usual consistency equation is enforced at
one scale, it will then no longer hold at any other. Put another way,
if the scalars are a perfect power-law, then the tensor spectrum
implied by the consistency relations is not (unless the spectral
indices are the same). Yet another way, the set of models generated by
imposing the consistency equation at one scale is a different set of
models from that obtained using another scale.  This problem is
further exacerbated if authors go on to include scalar spectral index
running, while perhaps still leaving the tensors as a
power-law. Before discussing the choice of scale, we should therefore
first fix this problem (while admitting that the difference may be too
small to be very important).

This is achieved by implementing the full inflationary consistency
equation hierarchy, as given explicitly in Ref.~\cite{CL}. As well as
the first consistency equation, this enforces that each
\emph{derivative} of the consistency equation also holds at a given
scale. When using a Taylor expansion to shift from one scale to
another, this hierarchy then ensures that the consistency equations
will still hold at the new scale (up to some level set by the
truncation of the hierarchy).

We note that these complications are needed only if one fits the
phenomenological parameters (amplitude, spectral index, running, etc)
from the data and then translates to inflationary observables. If
instead one fits the slow-roll parameters directly
(e.g.~Ref.~\cite{GL,finelli}) or via flow equations
\cite{flow,KKMR,PE} then the consistency equation hierarchy is
automatically enforced.

We consider a parametrization of the scalar and tensor 
perturbations as follows
\begin{eqnarray}
A_{\rm S}^2 (k)&\propto& (k/k_{*})^{(n_{\rm S}-1)+(dn_{\rm S}/d\ln
k) \, \ln k/k_*} \\
A_{\rm T}^2 (k)&\propto& (k/k_{*})^{n_{\rm T}+(dn_{\rm T}/d\ln
k) \, \ln k/k_*}\,,
\end{eqnarray}
the constants of proportionality being the amplitude of the
perturbations at scale $k_{*}$. The tensor-to-scalar ratio is defined
by $r(k)\equiv 16 A_{{\rm T}}^2(k)/A_{{\rm S}}^2(k)$, and the tensor
spectral index is determined via the first consistency equation.

In order that the first consistency equation is enforced at all scales
(to linear order in $\Delta \ln k$), we need to implement the second
consistency equation to fix the tensor running, which is not a genuine
new degree of freedom. This second equation is given by \cite{LLKCBA,CL}
\begin{equation}\label{2nd Cons}
\frac{dn_{\rm T}}{d \ln k}=n_{\rm T}\left[ n_{\rm T}-(n_{\rm
S}-1)\right]\,.
\end{equation}
We enforce this when carrying out our data-fitting.

One could further enforce higher consistency equations, so that for
instance the second consistency equation also is preserved under
change of scales. However current data quality is a long way from the
point where doing so would make any practical difference, since the
tensors are potentially observable only over a limited range of
scales.

We use the Monte Carlo Markov Chain (MCMC) technique to explore the
parameter space, using the CosmoMC package \cite{CosmoMC}.  We
consider a $\Lambda$CDM model in a flat universe and take $k_*=0.05 \,
{\rm Mpc}^{-1}$ as the scale where all power spectrum parameters are
defined when fitting to data. We vary up to eight parameters
\begin{equation}
\Omega_{{\rm b}} h^2, \Omega_{{\rm dm}} h^2, \theta, \tau,
n_{\rm S}(k_*), r(k_*), 
\ln [ 10^{10} A_{\rm S}(k_*)], \left. \frac{dn_{\rm
  S}}{d\ln k}\right|_{k_*} 
 \nonumber
\end{equation}
where $\Omega_{{\rm b}} h^2$ and $\Omega_{{\rm dm}} h^2$ are the
physical baryon and dark matter densities, $\theta$ is the ratio of
the sound horizon to the angular-diameter distance, $\tau$ is the
optical depth, and the remaining parameters specify the power spectra.
We apply a set of uniform priors:
\begin{center}
\begin{tabular}{cc}
  $0.005<\Omega_{{\rm b}} h^2 <0.1$ \, & $0.01<\Omega_{{\rm dm}}
  h^2<0.99\,$ \\
  $0.5<\theta<10\,$& $0.01<\tau<0.8 \,$ \\
  $0.5<n_{\rm S}<1.5\, $&$ 2.7<\log ( 10^{10} A_{\rm S})<4\,$\\
  $0<r<2\, $&$ -0.2<dn_{\rm S}/d\ln k <0.2\,$
\end{tabular}
\end{center}
Until Section~\ref{s:alldata}, our constraints are from WMAP3 data
alone.

\section{Choice of scale: models with scalar running}

We first consider models which allow running of the scalar spectral
index, which we will see is the case where the choice of scale is most
important. For comparison, models without running are studied in the
next section.

\subsection{Tilt and running}

The simplest combination of observables to consider is the tilt
and running of the scalars. Observational implications of this
were first discussed in Ref.~\cite{CGL}, which forecasted CMB
constraints from the Planck satellite on running spectral index
models. The paper pointed out that there would be a scale at which
the uncertainties on tilt and running would become uncorrelated,
and that (at least in a gaussian approximation) on that scale the
uncertainty in $n$ would recover its value for the case of no
running.\footnote{This observation was actually credited to Daniel
Eisenstein, who was not an author of that paper.} This could be
spoiled by degeneracies with other parameters, but at Planck
accuracy appears not to be \cite{CGL}.

\begin{figure}[t]
\includegraphics[width=8 cm]{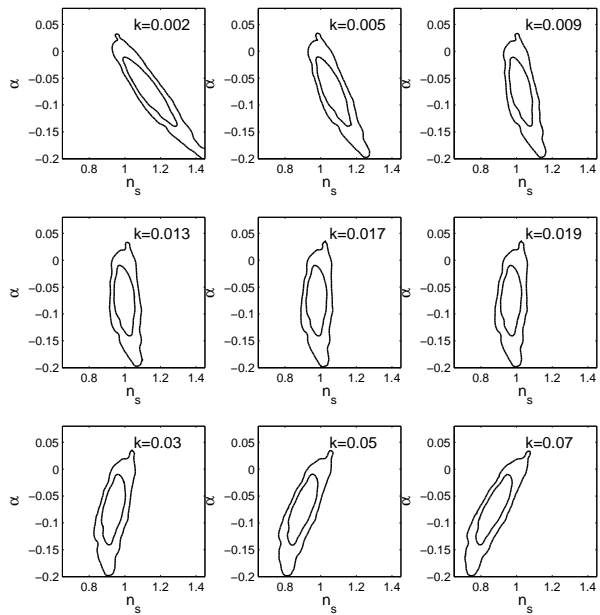}
\caption{Constraints in the $n_{\rm S}$--$\alpha$ plane (where $\alpha
= dn_{\rm S}/d\ln k$) at several scales. $k = 0.017 \, {\rm
Mpc}^{-1}$ is the decorrelation scale for these
parameters. \label{n_run}}
\end{figure}

Anyway, we wish to find the scale at which the tilt and running
decorrelate for actual current data. To do this we take the
distribution of these two variables as given by the MCMC analysis,
which specifies quantities at $k_*=0.05 \, {\rm Mpc}^{-1}$.  We then
fit the chain elements with a linear relation, $n_{\rm S}= A + B \,
dn_{\rm S}/d\ln k$, and by inserting into the expression 
\begin{equation}\label{tilt}
n_{\rm S}(k)=n_{\rm S}(k_*)+ \frac {dn_{\rm
S}}{d\ln k} \ln \frac{k}{k_*}\,,
\end{equation}
we arrive at a condition for the difference in scale which
decorrelates $n_{\rm S}$ and $dn_{\rm S}/d\ln k$: $B=-\ln k/k_*$.
This scale turns out to be $k=0.017 \, {\rm Mpc}^{-1}$. Then we use
Eq.~(\ref{tilt}) to convert the distribution at scale $k_*$ to the one
at scale $k=0.017 \, {\rm Mpc}^{-1}$ to obtain the decorrelated
$n_{\rm S}$ and $dn_{\rm S}/d\ln k$.  More generally, we can explore
the constraints at other scales via the same formalism. The
constraints at a set of different scales, including the WMAP scale and
the decorrelation scale, are shown in Fig.~\ref{n_run}.

In this particular case (but not those that follow), the
transformation between parameters induced by the scale change has unit
Jacobian, which means that the 2D contour areas are preserved.
However it is clear from Fig.~\ref{n_run} that the marginalized
uncertainty on $n_{{\rm S}}$ at the decorrelation scale is
significantly smaller.  The WMAP choice, $0.002 \, {\rm Mpc}^{-1}$,
gives a significantly-angled constraint area and is clearly to be
avoided if useful marginalized constraints on $n_{{\rm S}}$ are to be
quoted. Unfortunately, the main WMAP3 results for models with running
are presented at this scale.

For our choice of parameters and dataset (WMAP3 alone), from separate
fits where running is not included we find the marginalized constraint
on $n_{\rm S}$ is $n_{\rm S}=0.993_{-0.030,-0.053}^{+0.029,+0.067}$
(at 68\% and 95\% confidence). With running, the marginalized
constraint at the decorrelation scale is $n_{\rm S} =
0.981_{-0.034,-0.063}^{+0.034,+0.067}$.  As anticipated, therefore,
when including running the shift in the best-fit $n_{\rm S}$ at the
optimized scale is negligible within the uncertainty. This is somewhat
trivial as it could have been chosen to match exactly by specific
choice of scale --- choosing $k=0.015 \, {\rm Mpc}^{-1}$ achieves
this. Much more importantly, we see that the \emph{uncertainty} on
$n_{{\rm S}}$ at the decorrelation scale is hardly increased when
running is included, whereas it is greatly increased at
e.g.~$0.002 \, {\rm Mpc}^{-1}$. The 1D marginalized constraints on all
parameters have minimum uncertainty at the decorrelation scale.

Incidentally, for the scalar running the marginalized constraints
we obtained are $dn_{\rm S}/d\ln
k=-0.075_{-0.043,-0.093}^{+0.041,+0.082}$, very similar to those
quoted by WMAP3 for models with running and tensors \cite{wmap3}.

\subsection{Tilt and the tensor--scalar ratio}

We now turn to other combinations of observables, relevant to
constraining inflation.

\begin{figure}
\includegraphics[width=8 cm]{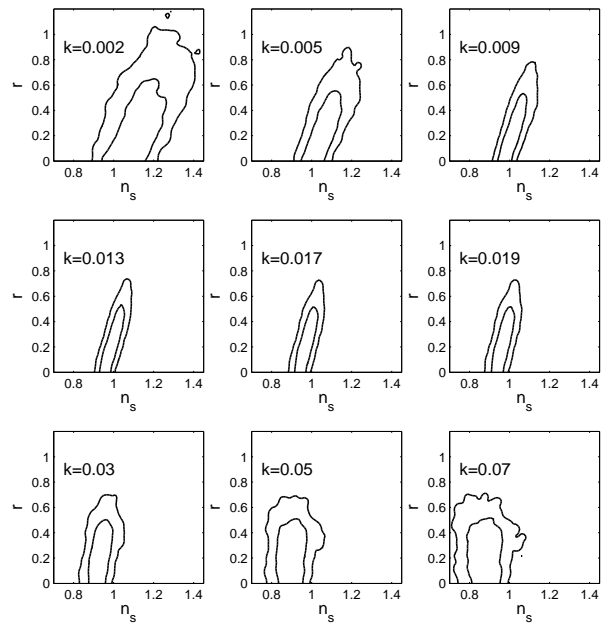}
\caption{Constraints on $n_{\rm S}$ versus $r$ at several
scales. \label{n_r}}
\end{figure}

To obtain $r$ at other scales we perform an expansion, to the order
considered, of the scalar and tensor amplitudes. The relation is
\begin{equation}\label{shift_r}
\frac{r(k)}{r(k_*)} =  \frac {1+n_{\rm T} \ln \frac{k}{k_*}+
\frac{1}{2}\left[n_{\rm T}^2+ \frac{dn_{\rm T}}{d\ln
k}\right] \ln^2 \frac{k}{k_*}}{1+(n_{\rm S}-1)
\ln \frac{k}{k_*} + \frac{1}{2}\left[(n_{\rm S}-1)^2+
\frac{dn_{\rm S}}{d\ln k}\right] \ln^2 \frac{k}{k_*}}\,.
\end{equation}
where all observables without an argument `(k)' are evaluated at $k_*
= 0.05 \, {\rm Mpc}^{-1}$, and where $dn_{\rm T}/d\ln k$ was set
according to the lowest-order version of the second consistency
equation, Eq.~(\ref{2nd Cons}).  Having expressions for $n_{\rm S}$
and $r$ at different scales, we can now choose several scales
and get the distribution of the two variables at each, shown in
Fig.~\ref{n_r}.

\begin{figure}[t]
\includegraphics[width = 5.8 cm]{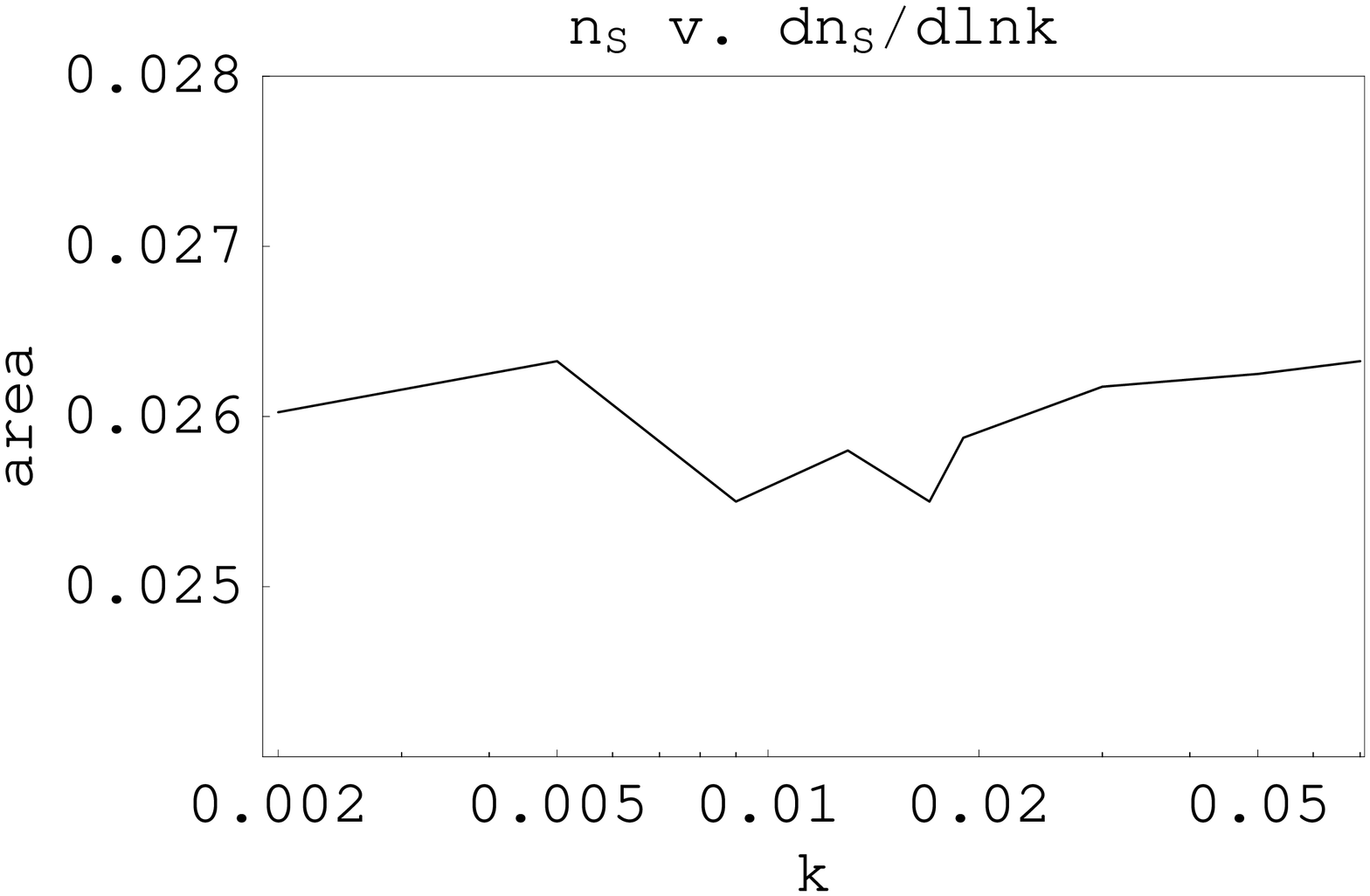}\\
\includegraphics[width = 5.8 cm]{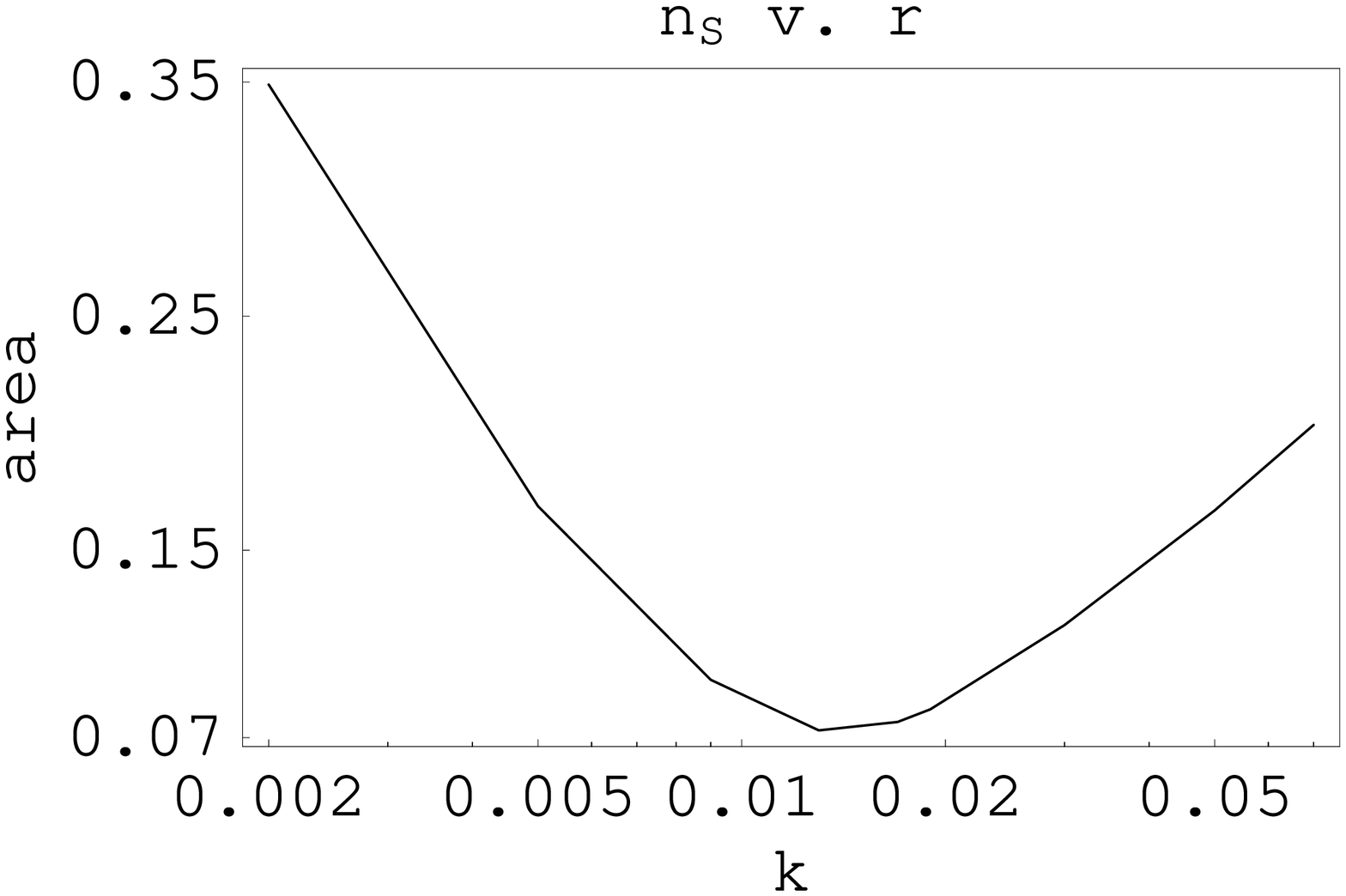}\\
\includegraphics[width = 5.8 cm]{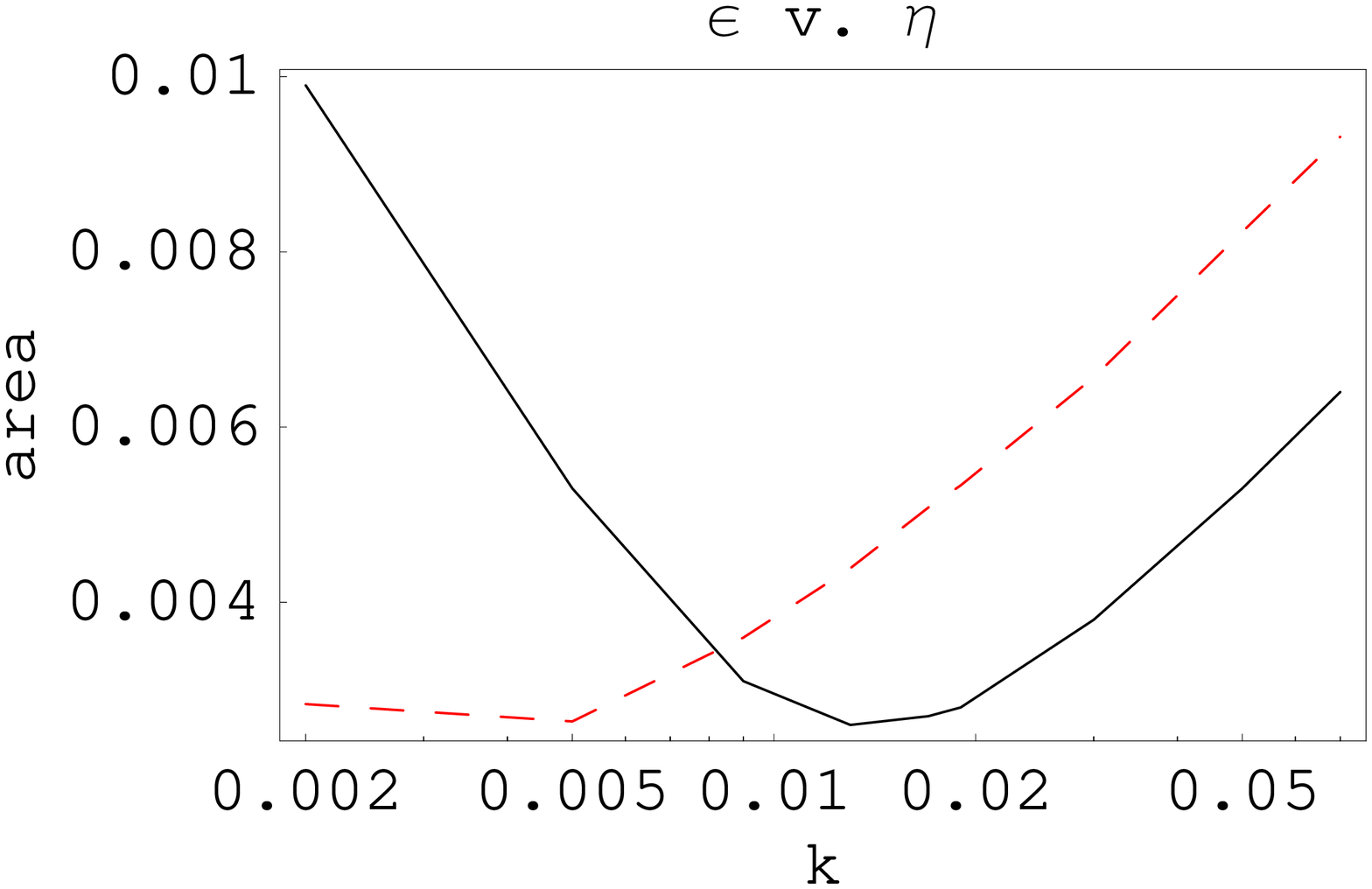}
\caption{Variation of parameter plane area with scale. For the
$\epsilon$--$\eta$ case both lowest (black, full) and next order (red,
dashed) are shown. For $n_{\rm S}$ versus running the area should be
independent of scale, and the variations indicate the noise level in
the area estimation.
\label{area-scale}}
\end{figure}

In this case the transformation alters the contour areas as well as
distorting them. The middle panel of Fig.~\ref{area-scale} shows the
areas enclosed by the 95\% confidence contour in the $n_{\rm S}$--$r$
plane at different scales.\footnote{These values were obtained by
taking the number of points in a $50\times 50$ grid that lie within
that contour. The number of grid points across each axis corresponds
also to the number of bins used to sample the distribution. We found
that accurate area estimation needed at least 50 bins, though such an
aggressive binning level leads to less smooth contours than are
usually seen.} The top panel shows the same for $n_{\rm S}$ and
running discussed in the previous subsection.  In the $n_{\rm S}$--$r$
plane the minimum area was near $k=0.017 \, {\rm Mpc}^{-1}$ as
expected (the precise value found was slightly smaller).  As inflation
model builders typically just look at these marginalized plots to
decide if their model is viable, it is clearly important to present
the constraints at a good scale. $0.002 \, {\rm Mpc}^{-1}$ is not a
good scale for this purpose, as has previously been stressed also in
Ref.~\cite{PE}.

\subsection{Inflationary slow-roll parameters: lowest order}

We now examine how the constraints on the first two slow-roll
parameters $\epsilon$ and $\eta$ are affected by scale change. We take
the usual definitions in terms of the potential \cite{LL92}
\begin{equation}
\epsilon =\frac{m_{{\rm
      Pl}}^2}{16\pi}\left(\frac{V'}{V}\right)^2 \quad ; \quad
\eta = \frac{m_{{\rm Pl}}^2}{8\pi}\, \frac{V''}{V}\,.
\end{equation}
The pivot scale $k_*$ corresponds to some particular scalar field
value $\phi_*$ (defined as the field value when $k_* = aH$ during
inflation), in the vicinity of which the scalar field potential is
being reconstructed. Shifting the pivot scale means expanding
about a different point on the potential.

We first concentrate on the constraints given at lowest order,
taking the expressions for the potential at this order by Lidsey
et al.~\cite{LLKCBA}:
\begin{eqnarray}
V(\phi)& \simeq &\frac{75 m_{\rm Pl}^4}{32}A_{\rm T}^2(k) \,,
\nonumber \\
V'(\phi)& \simeq &-\frac{75 \sqrt{\pi}}{8}m_{\rm Pl}^3\frac{A_{\rm
T}^3(k)}{A_{\rm S}(k)}\,,\\
V''(\phi)& \simeq &\frac{25\pi}{4}m_{\rm Pl}^2 A_{\rm
T}^2(k)\left[9\frac{A_{\rm T}^2(k)}{A_{\rm
S}^2(k)}-\frac{3}{2}[1-n_{\rm S}(k)]
  \right]\,, \nonumber
\end{eqnarray}
(where without loss of generality we take $\phi$ to increase in
time). From these the first two slow-roll parameters are expressed in
terms of the observables, to lowest order, by
\begin{equation}
\label{epsilon_eta_low} \epsilon \simeq \frac{r}{16} \quad ; \quad
\eta \simeq \frac{3}{16}r-\frac{1}{2}(1-n_{\rm S}) \,.
\end{equation}
Shifting the scale of the observables shifts the location on the
potential, and at lowest-order the constraints on $\epsilon$ and
$\eta$ then become independent of the running at that scale (which
could be used to determine a third slow-roll parameter $\xi \equiv
m_{{\rm Pl}}^2/8\pi \sqrt{ V' V'''/ V^2}$).

\begin{figure}[t]
\includegraphics[width=8 cm]{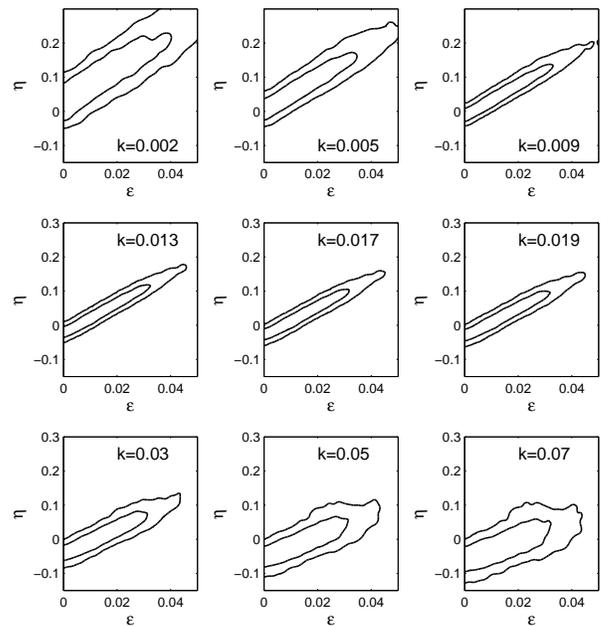}
\caption{Constraints on $\epsilon$ versus $\eta$, at
lowest order, evaluated at several scales.\label{eelow}}
\end{figure}

The results are presented in Fig~\ref{eelow}, and again show
strong variation of the allowed parameter area with choice of
scale, as indicated in Fig.~\ref{area-scale}.

\subsection{Inflationary slow-roll parameters: next order}

Now we can take the expressions for the potential to next order
\cite{SL}, also given by Lidsey et al.~\cite{LLKCBA}:
\begin{eqnarray}
V(\phi)&\simeq&\frac{75 m_{\rm Pl}^4}{32}A_{\rm
T}^2\left[1+\left(\frac{5}{3}+2C\right)\frac{A_{\rm
T}^2}{A_{\rm S}^2}\right] \,,\nonumber\\
V'(\phi)&\simeq&-\frac{75 \sqrt{\pi}}{8}m_{\rm Pl}^3\frac{A_{\rm
T}^3}{A_{\rm S}} \left[1-0.85\frac{A_{\rm
T}^2}{A_{\rm S}^2}+0.53(1-n_{\rm S})\right]\,,\nonumber\\
V''(\phi)&\simeq&\frac{25\pi}{4}m_{\rm Pl}^2 A_{\rm
T}^2\Big\{9\frac{A_{\rm T}^2}{A_{\rm S}^2}-\frac{3}{2}(1-n_{\rm
S})\\ && +\Big[(36C+2)\frac{A_{\rm T}^4}{A_{\rm S}^4}
-\frac{1}{4}(1-n_{\rm S})^2- \nonumber \\
&& \quad (12C-6)\frac{A_{\rm T}^2}{A_{\rm S}^2}(1-n_{\rm
S})-\frac{1}{2}(3C-1)\frac{dn_{\rm S}}{d\ln k}\Big]\Big\}\,, \nonumber
\end{eqnarray}
where $C=-2+\ln 2+\gamma\simeq-0.73$, $\gamma$ is the
Euler--Mascheroni constant, and again the $\phi$ value corresponds to
horizon crossing of the scale at which the constraints are being
imposed.

\begin{figure}[t]
\includegraphics[width=8 cm]{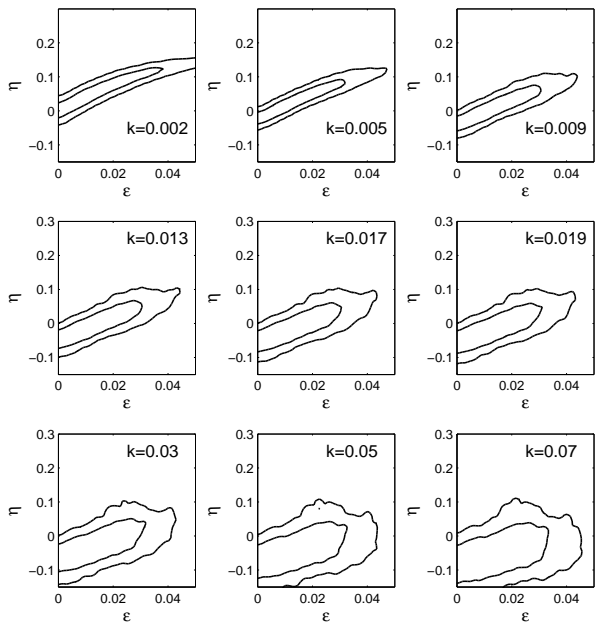}
\caption{Constraints on $\epsilon$ versus $\eta$, to next order, at
several scales. \label{eenext}} 
\end{figure}

With these next-order expressions for the potential, $\epsilon$ and
$\eta$ are 
\begin{eqnarray}\label{epsilon_eta_next}
\epsilon&=&\frac{r}{16} \, \frac{1-0.85\, r/16+0.53(1-n_{\rm
S})}{1+0.21\,r/16}\\
\eta&=&\frac{1}{3}\frac{1}{1+0.21 \, r/16}
\Big\{\frac{9}{16}r-\frac{3}{2}(1-n_{\rm S}) \\
&& +(36C+2)\left(\frac{r}{16}\right)^2 -\frac{1}{4}(1-n_{\rm
S})^2\nonumber \\
&& -(12C-6)\frac{r}{16}(1-n_{\rm
S}) -\frac{1}{2}(3C-1)\frac{dn_{\rm S}}{d\ln k}\Big\} \,. \nonumber
\end{eqnarray}
The second parameter now depends on the running. The running term has
a coefficient of about one half, and given how weakly running is
constrained this term has a significant impact on the constraints.

The constraints at each scale are presented in Fig~\ref{eenext}.  The
picture here is rather different, with the area changing much more
slowly as $k$ is decreased, and the minimum area being at a much
smaller $k$. This is because for typical models the next-order
correction from the running happens to be comparable to the change in
the lowest-order expression for $\eta$ coming from the changing
$n_{{\rm S}}$, also induced by the running, as the scale
changes. These terms approximately cancel going to smaller $k$, i.e.\
the constraints change less when simultaneously reducing $k$ and
introducing next-order corrections than they would if only one of
these were done. This is just a coincidence (and not much of a
coincidence at that, since partial cancellation would have to happen
as $k$ was changed in one or other direction) of no great
significance, and will go away when in future running is better
constrained.

\begin{figure}[t]
\includegraphics[width= 8cm]{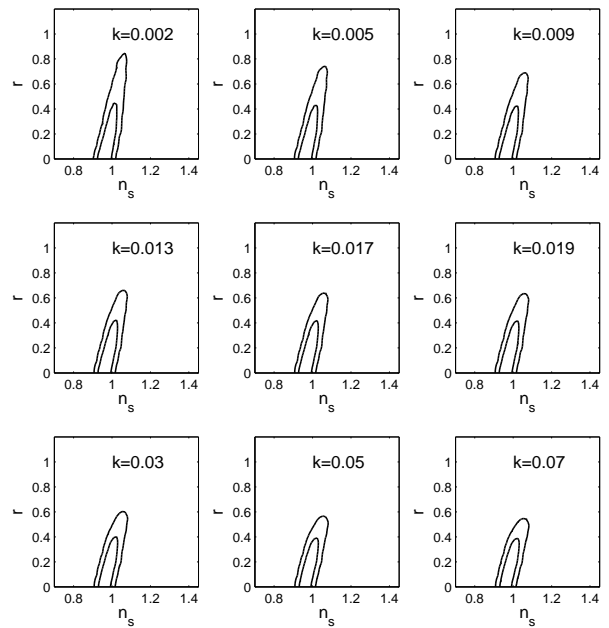}
\caption{Constraints on $n_{\rm S}$ and $r$ when no scalar running is
present.\label{n_r_no_run}}
\end{figure}

\section{Choice of scale: models with no scalar running}

For comparison, we now take a look at models where no running of the
scalar index is allowed.\footnote{We still keep the tensor running in
the analysis, however. It is not an additional degree of freedom, its
inclusion ensuring the validity of the first consistency equation at
all scales.} In this case the variation in the constraints with scale
is much less, as for instance is seen in Fig.~\ref{n_r_no_run} showing
the $n_{\rm S}$--$r$ plane. Indeed in this case we find that
minimization of the area is not only unnecessary, but can actually be
misleading, because parameters such as $r$ can appear to be well
constrained even on scales where there is no meaningful data. The
reason for this is that the restrictive class of models under
consideration force the spectra to behave in a particular way as they
are extrapolated away from the region where the bulk of the data lie,
i.e.~such constraints contain significant prior information as well as
data information. This is also true to some extent for constraints on
$r$ in the running case studied earlier.

\begin{figure}[t]
\includegraphics[width= 6.8cm]{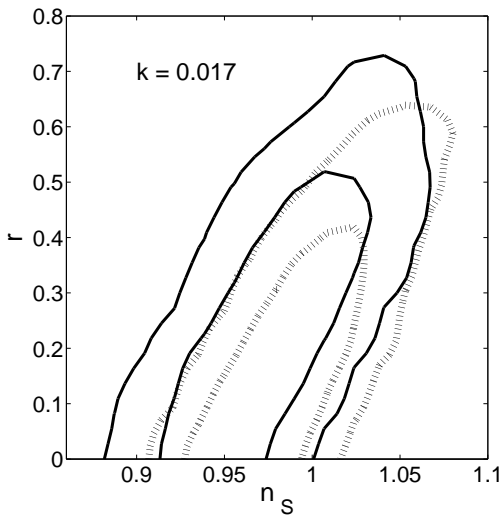}
\caption{Comparison of constraints in the $n_{\rm S}$--$r$ plane at
the optimal scale with no running (dotted contours) and when running
is included (full contours). The area enclosed by the 95\% contour
increases by around 20\% when running is included. \label{overlay}}
\end{figure}

Nevertheless, it is now interesting to compare the running and
no-running constraints. In the WMAP3 analysis the impression, from
comparison of the top-left panels of Figs.~12 and 14 of
Ref.~\cite{wmap3}, is of a huge deterioration in the constraints in
the $n_{\rm S}$--$r$ plane once running is included. The same is seen
in Fig.~1 of Ref.~\cite{KKMR}. However we now see that this is an
artifact of the choice of scale where the constraints are
portrayed. At the optimal scale there is some deterioration, due to
parameter degeneracy, but the area increase within the 95\% contour is
only by about 20\% as seen in Fig.~\ref{overlay}, not by a factor of
five as at $k=0.002 \, {\rm Mpc}^{-1}$. Consequently, inclusion of
running leads only to a moderate deterioration in constraints on
$\epsilon$ and $\eta$.

\section{Including more data}

\label{s:alldata}

We explore the robustness of our results by carrying out the same
analysis for a broader compilation of data, now including
shorter-scale CMB experiments and galaxy correlation data from ACBAR
\cite{ACBAR}, CBI \cite{CBI}, VSA \cite{VSA}, Boomerang \cite{boom},
SDSS \cite{SDSS}, and 2dFGRS \cite{2df}.

Everything goes through as before. We find that the decorrelation
scale of $n_{\rm S}$ and running is $0.016 \, {\rm Mpc}^{-1}$, which
is not significantly different from WMAP3 alone. Though in general one
would expect the decorrelation scale to change with dataset, in this
case the WMAP3 data are powerful enough that a shift is not seen.

\begin{figure}[t]
\includegraphics[width=8 cm]{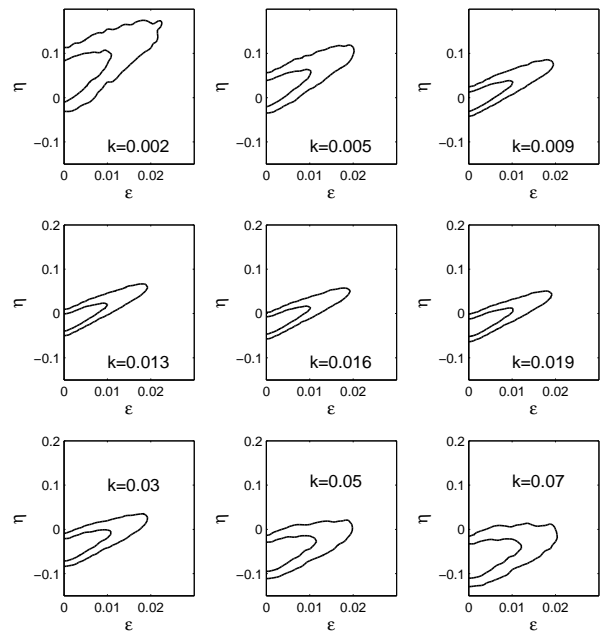}
\caption{As Fig.~\ref{eelow}, but now with the full dataset
compilation. Note the modified axis ranges. \label{f:alleelow}}
\end{figure}

\begin{figure}[t]
\includegraphics[width= 6.8 cm]{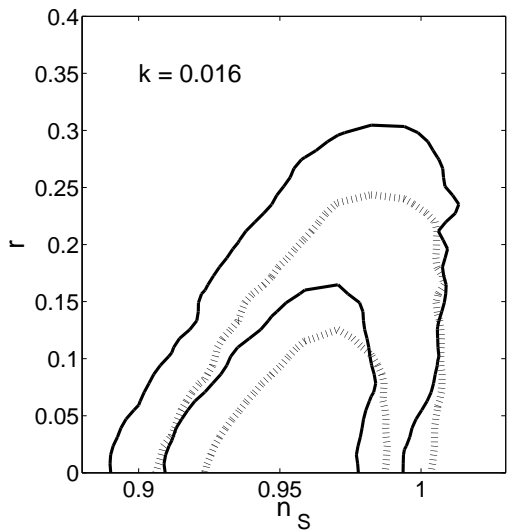}
\caption{As Fig.~\ref{overlay}, but now with the full dataset
compilation.  Note the modified axis ranges. The area enclosed by the
95\% contour increases by around 30\% when running is included.
\label{f:alloverlay}}
\end{figure}

The constraints, particularly on $r$ and hence $\epsilon$, do tighten
significantly with the extra data, as is clear also in previous
analyses including Ref.~\cite{wmap3}. As an illustration of the
results we obtained in this case, we show the array of constraints on
the lowest-order $\epsilon$ and $\eta$ at different scales,
Fig.~\ref{f:alleelow}, and the overlay of contours in the $n_{\rm
  S}$--$r$ plane at the optimal scale, with and without running, in
Fig.~\ref{f:alloverlay}. 

\section{Conclusions}

We have investigated the issue of choice of scale in presenting
marginalized parameter constraints. While we have focussed on WMAP
constraints applied to inflationary models, the same considerations
apply much more widely. For example, in constraining density
perturbations using galaxy clusters, commonly the parameter
$\sigma_8$, being the normalization of density perturbations smoothed
on the scale $8 h^{-1} \, {\rm Mpc}^{-1}$, is quoted. However
typically the normalization is best determined at a somewhat larger
scale than $8h^{-1} \, {\rm Mpc}$, and marginalizing over parameters
such as $\Omega_0$ to quote constraints on $\sigma_8$ can
unnecessarily increase the statistical uncertainty on the
normalization.

In the inflationary context, choosing an optimal scale is important
primarily in models where large running is allowed. We found that an
appropriate scale is the one which decorrelates estimates of $n_{\rm
S}$ and running, which for WMAP3 is $0.017 \, {\rm Mpc}^{-1}$.  This
criterion can be used to define such a scale for any dataset
compilation, and we found that the scale shifts hardly at all when
other available data are added to WMAP3.  The optimal scale may also
have some modest dependence on the choice of model parameters varied
in a fit, for instance if non-negligible neutrino masses were included.
One might even wonder
whether it might be best to choose different scales for different
observables, as the scalars and tensors are best constrained on quite
different length scales, but we have not attempted this here.

We have shown that the marginalized constraints on $n_{\rm S}$ and
$r$, or on $\epsilon$ and $\eta$, depend significantly on the choice
of scale in the presence of running. By choosing the optimal scale, we
find that constraints on those parameters are only mildly degraded by
the inclusion of running as a parameter, in contrast to the impression
given if constraints are quoted at a non-optimal scale such as $0.002
\, {\rm Mpc}^{-1}$.


\begin{acknowledgments}
M.C.\ was supported by FCT (Portugal), and A.R.L.\ and P.M.\ by PPARC
(UK).  M.C.\ thanks Bruce Bassett for hospitality during this
work. A.R.L.\ thanks the Institute for Astronomy, University of
Hawai`i, for hospitality during this work. We thank Richard Easther
and Hiranya Peiris for helpful discussions and comments.
\end{acknowledgments}




\begin{thebibliography}{}
\bibitem{wmap3} D. N. Spergel et al.~[WMAP collaboration],
  astro-ph/0603449.
\bibitem{KMV} H. Kurki-Suonio, V. Muhonen, and J. Valiviita,
  Phys. Rev.  D{\bf 71}, 063005 (2005), astro-ph/0412439.
\bibitem{finelli} F. Finelli, M. Rianna, and N. Mandolesi, JCAP {\bf
  0612}, 006 (2006), astro-ph/0608277.
\bibitem{CosmoMC} A. Lewis and S. L. Bridle, Phys. Rev. D{\bf 66},
  103511 (2002), astro-ph/0205436.
\bibitem{LLMP} A. R. Liddle, D. Parkinson, S. M. Leach, and P. Mukherjee, 
  Phys. Rev. D{\bf 74}, 083512 (2006), astro-ph/0607275.
\bibitem{PE} H. Peiris and R. Easther, JCAP {\bf 0610}, 017 (2006),
  astro-ph/0609003. 
\bibitem{run} J. E. Lidsey and R. Tavakol, Phys. Lett. B{\bf 575}, 157
  (2003), astro-ph/0304113; D. J. H. Chung, G. Shiu, and M. Trodden,
  Phys. Rev. D{\bf 68}, 063501 (2003), astro-ph/0305193; J. M. Cline and
  L. Hoi, JCAP {\bf 0606}, 007 (2006), astro-ph/0603403; C. Pahud,
  A. R. Liddle, P. Mukherjee,  and D. Parkinson, astro-ph/0701481.
\bibitem{EP} R. Easther and H. Peiris, JCAP {\bf 0609}, 010 (2006),
astro-ph/0604214. 
\bibitem{KKMR} W. H. Kinney, E. W. Kolb, A. Melchiorri, and
  A. Riotto, Phys. Rev. D{\bf 74}, 023502 (2006), astro-ph/0605338.
\bibitem{LLKCBA} J. E. Lidsey, A. R. Liddle, E. W. Kolb, E. J. Copeland,
    T. Barreiro, and M. Abney, Rev. Mod. Phys. {\bf 69}, 373 (1997),
    astro-ph/9508078.
\bibitem{CL} M. Cort\^es and A. R. Liddle, Phys. Rev. D{73},
  083523 (2006), astro-ph/0603016.
\bibitem{GL} I. J. Grivell and A. R. Liddle, Phys. Rev. D{\bf 61},
081301 (2000), astro-ph/9906327; S. M. Leach, A. R. Liddle, J. Martin,
and D. J. Schwarz, Phys. Rev. D{\bf 66}, 023515 (2002),
astro-ph/0202094; J. Martin and C. Ringeval, JCAP {\bf 0501}, 007
(2005), astro-ph/0605367.
\bibitem{flow} M. B. Hoffman and M. S. Turner, Phys. Rev. D{\bf 64},
  023506 (2001), astro-ph/0006321; W. H. Kinney,
  Phys. Rev. D{\bf 66}, 083508 (2002), astro-ph/0206032;
  H. Peiris and R. Easther, JCAP {\bf 0607}, 002 (2006),
  astro-ph/0603587.
\bibitem{CGL} E. J. Copeland, I. J. Grivell, and A. R. Liddle,
  Mon. Not. Roy. Astron. Soc. {\bf 298}, 1233 (1998),
  astro-ph/9712028.
\bibitem{LL92} A. R. Liddle and D. H. Lyth, Phys. Lett. B{\bf 291},
  391 (1992), astro-ph/9208007.
\bibitem{SL} E. D. Stewart and D. H. Lyth, Phys. Lett. B{\bf 302},
    171 (1993), gr-qc/9302019.
\bibitem{ACBAR} C. L. Kuo et al., Astrophys. J. {\bf 600}, 32 (2004),
	astro-ph/0212289.
\bibitem{CBI} T. J. Pearson et al., Astrophys. J. {\bf 591}, 556 
	(2003), astro-ph/0205388.  
\bibitem{VSA} C. Dickinson et al., Mon. Not. Roy. Astron. Soc. {\bf 
	353}, 732 (2004), astro-ph/0402498.  
\bibitem{boom} W. C. Jones et al., Astrophys. J. {\bf 647}, 823
       (2006), astro-ph/0507494. 
\bibitem{SDSS} M. Tegmark et al., Astrophys. J. {\bf 606}, 702 (2004),
	astro-ph/0310725. 
\bibitem{2df} W. Percival et al., Mon. Not. Roy. Astron. Soc. {\bf 
	327}, 1297 (2001), astro-ph/0105252. 
\end{thebibliography}
\end{document}